\pdfoutput=1
\RequirePackage{ifpdf}
\ifpdf % We are running pdfTeX in pdf mode
\documentclass[pdftex]{sigma}
\else
\documentclass{sigma}
\fi

\numberwithin{equation}{section}

\begin{document}

\newcommand{\arXivNumber}{1502.03673}

\allowdisplaybreaks

\renewcommand{\thefootnote}{$\star$}

\renewcommand{\PaperNumber}{047}

\FirstPageHeading

\ShortArticleName{Higher Order Deformations of Complex Structures}

\ArticleName{Higher Order Deformations of Complex Structures\footnote{This paper is a~contribution to the Special Issue on Exact Solvability and Symmetry Avatars
in honour of Luc Vinet.
The full collection is available at
\href{http://www.emis.de/journals/SIGMA/ESSA2014.html}{http://www.emis.de/journals/SIGMA/ESSA2014.html}}}

\Author{Eric D'HOKER~$^\dag$ and Duong H. PHONG~$^\ddag$}

\AuthorNameForHeading{E.~D'Hoker and D.H.~Phong}

\Address{$^\dag$~Department of Physics and Astronomy, University of California, Los Angeles 90024, USA}
\EmailD{\href{mailto:dhoker@physics.ucla.edu}{dhoker@physics.ucla.edu}}

\Address{$^\ddag$~Department of Mathematics, Columbia University, New York, NY 10027, USA}
\EmailD{\href{mailto:phong@math.columbia.edu}{phong@math.columbia.edu}}

\ArticleDates{Received March 27, 2015, in f\/inal form June 15, 2015; Published online June 23, 2015}

\Abstract{Deformations of complex structures by f\/inite Beltrami dif\/ferentials are consi\-de\-red on general Riemann surfaces. Exact formulas to any f\/ixed order are derived for the corresponding deformations of the period matrix, Green's functions, and correlation functions in conformal f\/ield theories with vanishing total central charge. The stress tensor is shown to give a simple representation of these deformations valid to all orders. Such deformation formulas naturally enter into the evaluation of superstring amplitudes at two-loop order with Ramond punctures, and at higher loop order, in the supergravity formulation of the RNS superstring.}


\Classification{32G05; 51M15; 51P05; 53Z05} 

\renewcommand{\thefootnote}{\arabic{footnote}}
\setcounter{footnote}{0}

\section{Introduction}

It is a fundamental property of the moduli space of Riemann surfaces that its tangent space is given by the space of Beltrami dif\/ferentials, modulo the range of the $\bar\partial$ operator on vector f\/ields. In general, without some additional structure, such as a connection, a~tangent vector to a~mani\-fold determines only an inf\/initesimal deformation on the manifold. However, in the case of moduli space, we can actually associate to each Beltrami dif\/ferential a f\/inite deformation of the underlying complex structure. It is the purpose of this note to work out in detail the properties of this f\/inite deformation, including explicit formulas to all orders in the Beltrami dif\/ferential for the resulting period matrix and Green functions. Similarly, it is well-known that inf\/initesimally, the deformations of the correlation functions of a conformal f\/ield theory can be obtained by a~single insertion of the stress tensor. We show that a simple exponentiated generalization of this formula remains valid to all orders in  a f\/inite deformation, assuming the vanishing of the total central charge.

The problem of higher order deformations of complex structures arises in superstring perturbation theory. There the period matrix of a super Riemann surface can receive corrections of higher order as the gravitino f\/ield is turned on. For the scattering of Neveu--Schwarz states in genus~$2$, deformations by a Beltrami dif\/ferential are needed only to f\/irst order  \cite{D'Hoker:2001nj,D'Hoker:2001zp,D'Hoker:2005jc} (see also~\cite{D'Hoker:2002gw} for a~review). This is no longer the case for genus $3$ or higher, or even for genus $2$, when the scattering of Ramond states is considered \cite{D'Hoker:2015kwa, Witten:2015hwa}. The results of the present note will play,  in particular, an essential role in the construction of string amplitudes with Ramond states at two-loop level. We shall report on these issues elsewhere.

\section{Finite deformations by Beltrami dif\/ferentials}

Let $\Sigma$ be a compact Riemann surface, and let $z$ denote a local holomorphic coordinate on the surface. Let $\mu=\mu(z)\,d\bar z\otimes \partial_z$ be a Beltrami dif\/ferential. We shall always assume that the reduced part $\mu_{\rm red}$ of $\mu$
satisf\/ies
$|\mu_{\rm red}|<1$, and that $\mu$ has even grading in a f\/initely-generated Grassmann algebra. This last situation is the one of interest in superstring perturbation theory, where furthermore $\mu$ is  actually a nilpotent element of f\/inite order of a Grassmann algebra, the order being related to the number of odd moduli in the problem.

We def\/ine a new, deformed,  complex structure on $\Sigma$ by the requirement that its local holomorphic coordinate $w$ satisfy the equation
\begin{gather*}
%\label{w}
(\partial_{\bar z}+\mu (z)\partial_z)w=0.
\end{gather*}
The well-known theorem of Beltrami asserts that a local holomorphic coordinate $w$ always exists, and that the partial derivative~$\partial_z w$ may be taken to be non-vanishing. We shall often distinguish the two complex structures by indicating their respective local complex coordinate, either~$z$ for the original complex structure  or $w$ for the deformed complex structure.
For a given complex structure, we decompose the cotangent bundle to $\Sigma$ into the canonical bundle~$K$ and its complex conjugate~$\bar K$, whose sections are referred to as forms of type~$(1,0)$ and~$(0,1)$ respectively. When several complex structures, say~$z$ and~$w$, are simultaneously at play we shall use the notations~$K_z$ and~$K_w$ for their respective cotangent bundles.

\subsection{Deformation of holomorphic 1-forms}

Consider now a holomorphic (1,0)-form $\omega$ with respect to the deformed complex structure~$w$, so that~$\omega$ is a holomorphic section of~$K_w$, and takes the form $\omega=\omega(w)dw$. The basic observation is that, expressed in terms of~$z$ coordinates, a holomorphic $(1,0)$-form with respect to the complex structure~$w$ must be of the form
\begin{gather}
\label{omega}
\omega=\varphi dz-(\mu\varphi) d\bar z,
\end{gather}
with the coef\/f\/icient function $\varphi$ satisfying the relation
\begin{gather}
\label{closed}
\partial_{\bar z}\varphi=-\partial_z(\mu\varphi).
\end{gather}
Indeed, since we have the equations
\begin{gather}
dw   =   (\partial_z w) \,dz+ (\partial_{\bar z} w) \,d\bar z = ( \partial_z w ) (dz-\mu(z) d\bar z),
\nonumber \\
\partial_{\bar w}  =  (\partial_{\bar w} z) \partial_z + (\partial_{\bar w} \bar z) \partial_{\bar z}
  = (\partial_{\bar w} \bar z)(\partial_{\bar z}+\mu(z)\partial_z),\label{2a4}
\end{gather}
it follows that $\omega$ can  be expressed in the form (\ref{omega}) by taking
$\varphi=(\partial_zw)\omega(w)$. Furthermore, the holomorphicity of a $(1,0)$-form is equivalent to the fact that it is closed, which can be implemented using  either set of coordinates,~$z$ or~$w$. Using the coordinate~$z$, we f\/ind that $d\omega=0$ is equivalent (\ref{closed}), giving the desired equality.

The general solution of  equation (\ref{closed}) is given by the following integral equation\footnote{Throughout, we shall use the conventions of \cite{D'Hoker:1988ta}, such as the normalization $d^2z = i dz \wedge d\bar z$.}
\begin{gather}
\label{closed1}
\varphi(z)
=
\psi(z) + {1\over 2\pi} \int _\Sigma d^2 z'   \partial_z\partial_{z'} \ln E(z,z') (\mu  \varphi ) (z'),
\end{gather}
where $\psi = \psi(z) dz$ is an arbitrary  holomorphic $(1,0)$-form and $E(z,z')$ is the prime form, both with respect to the complex structure $z$. Successively iterating (\ref{closed1}) gives an expression for $\varphi$ to an arbitrary order  in $\mu$. This iteration process terminates after a f\/inite number of iterations  to give the exact solution when $\mu$ is a nilpotent element valued in a f\/initely-generated Grassmann algebra, which is the case  in superstring perturbation theory.

\subsection{Deformation of the period matrix}

We can evaluate now the deformation of the period matrix of the Riemann surface $\Sigma$ to arbitrarily high order in $\mu$. For this, we f\/ix a canonical homology basis $A_I$, $B_I$, $1\leq I\leq h$, $\#(A_I\cap B_J)=\delta_{IJ}$, $\#(A_I\cap A_J)=0$, $\#(B_I\cap B_J)=0$, where $h$ is the genus of $\Sigma$. Let $\psi_I(z)dz$ be the basis of holomorphic 1-forms with respect to the complex structure~$z$ which is dual to the cycles~$A_I$, and let $\Omega_{IJ}$ be the matrix of their~$B_J$ periods
\begin{gather*}
\oint_{A_J}  \psi_I(z)   dz =\delta_{IJ},
\qquad
\oint_{B_J} \psi_I(z)   dz=\Omega_{IJ}.
\end{gather*}
Corresponding to the basis~$\psi_I$ is a basis $\omega_I$ of holomorphic $(1,0)$-forms with respect to the $w$ structure, with $\omega_I=\varphi_I(z) dz-(\mu\varphi_I)(z) d\bar z$ and $\varphi_I(z)$ given by (\ref{closed1})
\begin{gather}
\label{closed2}
\varphi_I(z)
=
\psi_I(z)
+
{1\over 2\pi}\int _\Sigma d^2z'   \partial_z\partial_{z'} \ln E(z,z') (\mu  \varphi_I) (z').
\end{gather}
Their periods around a closed cycle $C$  are given by
\begin{gather}
\label{C}
\oint_C \omega_I(w) dw
=
\oint_C \big ( \varphi_I(z) dz-(\mu\varphi_I)(z)\,d\bar z \big ) .
\end{gather}
To express these periods in terms of the complex structure~$z$ and the Beltrami dif\/ferential $\mu$, we substitute in this formula the right hand side of (\ref{closed2}) for $\varphi_I$. But due to the fact that the integral of $\partial_z\partial_{z'} \ln E(z,z')$ is only conditionally convergent, the interchange of the order of integrations in~$z$ and~$z'$ is found to require a correction term, and is given by the following formula proved in~\cite{D'Hoker:1989ai}
\begin{gather}
{1 \over 2 \pi} \oint_C dz \left ( \int _\Sigma d^2z'  \partial_z\partial_{z'} \ln E(z,z') (\mu \varphi_I) (z')  \right )
\nonumber\\
\qquad{} =
{1 \over 2 \pi}  \int _\Sigma d^2z'   \left ( \oint_C dz    \partial_z\partial_{z'} \ln E(z,z') \right ) (\mu\varphi_I)(z')
+
\oint_C(\mu\varphi_I)(z) d\bar z.\label{C1}
\end{gather}
Combining (\ref{C}) and (\ref{C1}), we see that the integral over the $(0,1)$-form $(\mu \varphi_I)d\bar z$
cancels out, and we are left with
\begin{gather*}
\oint_C \omega_I(w) dw
=
\oint_C\psi_I(z) dz
+
{1\over 2\pi}\int _\Sigma d^2z' \left ( \oint_C \partial_z\partial_{z'} \ln E(z,z')  dz \right ) (\mu\varphi_I)(z')  .
\end{gather*}
Now the monodromy of the prime form $E(z,w)$ is trivial around $A_I$-cycles, and around $B_I$-cycles is  given by
\begin{gather*}
E(z+B_J,z')=E(z,z')   \exp \left ( -\pi i\Omega_{JJ}+2\pi i\int_z^{z'}\omega_J \right ).
\end{gather*}
Letting $C=A_J$ or $C=B_J$ in the preceding formula, we f\/ind
\begin{gather*}
\oint_{A_J}\omega_I(w) dw=\delta_{IJ},
\qquad
\oint_{B_J}\omega_I(w) dw
=
\Omega_{IJ}
+
i\int_\Sigma  d^2z  \,  \omega_J(z)\mu(z)\varphi_I(z),
\end{gather*}
where $\varphi_I(z')$ may be obtained by successive iterations of (\ref{closed1}).

Thus, the forms $\omega_I$  precisely provide the basis of holomorphic one-forms with respect to the deformed complex structure $w$ which is dual to the cycles $A_I$. Their $B_J$-periods provide the period matrix $\Omega_{IJ}(\mu)$ of the  complex structure $w$ in terms of the period matrix $\Omega _{IJ} = \Omega _{IJ}(0)$ of the original complex structure $z$. Repeated iteration of the formula (\ref{closed2}) for $\varphi_I$ produces $\Omega_{IJ}(\mu)$ to an arbitrary f\/ixed order  in a power expansion of $\mu$. For example, to order ${\cal O}(\mu^3)$, we obtain
\begin{gather*}
\Omega_{IJ}(\mu)
 = \Omega_{IJ}+i\int_\Sigma d^2z \, \omega_J(z)\mu(z)\omega_I(z)
\\
\hphantom{\Omega_{IJ}(\mu)=}{}
+
{i\over 2\pi}\int_\Sigma d^2 z \int_\Sigma d^2 z' \, \omega_J(z) \mu(z) \partial_{z} \partial_{z'} \ln E(z,z') \mu(z')\omega_I(z' )
+
{\cal O}\big(\mu^3\big).
\end{gather*}
For f\/ixed $z$, the integral over~$z'$ is conditionally convergent, but may naturally be def\/ined uniquely
by requiring that the integral over~$z'$ of the singularity $(z-z')^{-2}$ over a small circular disc centered at~$z$ must vanish by angular integration.

\section{Deformations of tensors and covariant derivatives}

Next, we shall consider  deformations of tensors of more general weight. It will be convenient to  denote by $T^{m,n}(\Sigma_z)$ the space of sections of $K_z^{\otimes m} \otimes \bar K_z ^{\otimes n}$, namely the space of tensors of type $\varphi (z) (dz)^m\otimes (d\bar z)^n$, where $K_z$ is the canonical bundle for complex structure $z$. Here, the subindex $z$ continues to indicate the complex structure.  Spinors may be included as well by  taking half-integer $m$ and $n$, and providing the additional data of a spin structure.

\subsection{Deformation of tensors}

Now consider a tensor $\varphi (z) (dz)^n$ belonging to $T^{(n,0)}(\Sigma _z)$ for given $n$. Then the correspondence $\omega\,dw\leftrightarrow \varphi \,dz$ used in the previous section for the special case $n=1$ generalizes to the following correspondences for tensors of type $\varphi (z) (dz)^n$
\begin{gather*}
\iota_{w\leftarrow z}\colon \  \varphi (z)   \to  \tilde \varphi (w) = \varphi (z) (\partial_w z)^n,
 \\
\iota_{z\leftarrow w}\colon \  \varphi (w)   \to   \tilde \varphi (z) = \varphi (w)   (\partial_z w)^n.
\end{gather*}
Intrinsically, the map $\iota_{z \leftarrow w}$ is proportional  to the projection from the space
$T^{n,0}(\Sigma_w)$ onto the space~$T^{n,0}(\Sigma_z)$, def\/ined by the direct sum decomposition
\begin{gather*}
T^{n,0}(\Sigma_w) = \bigoplus_{k=0}^n T^{n-k,k}(\Sigma_z)
\end{gather*}
and keeping only the top component $T^{n,0}(\Sigma_z)$. This decomposition may be carried out explicitly, with the help of~(\ref{2a4})
\begin{gather*}
\varphi (w) (dw)^n =
\sum _{k=0}^n { n ! \over k! (n-k)!} (-\mu)^k \varphi(w) (\partial_zw)^n (dz)^{n-k} \otimes (d\bar z)^k.
\end{gather*}
The  map $\iota _{z \leftarrow w}$ corresponds to retaining only the term $k=0$.
Similarly, the  inverse map $\iota_{w \leftarrow z}$ is def\/ined by the analogous projection from $T^{n,0}(\Sigma_z)$ onto~$T^{n,0}(\Sigma_w)$. We note that the maps~$\iota_{z\leftarrow w}$ and $\iota_{w\leftarrow z}$ restricted to the spaces $T^{n,0}(\Sigma_w)$ onto $T^{n,0}(\Sigma_z)$ are not inverses of each other. A~simple calculation gives their composition as follows
\begin{gather}
\label{3a3}
\iota_{z\leftarrow w}   \iota_{w\leftarrow z}=(\partial_zw)^n(\partial_w z)^n={1\over (1-\mu\bar\mu)^n}.
\end{gather}
Thus, the maps are not properly projections but are proportional to  projections.

\subsection{Deformation of covariant derivatives}

In the treatment of deformations of one-forms in the previous section, it suf\/f\/iced to use the de Rham exterior derivative. However, for tensors of higher rank we shall need covariant derivatives. So we introduce now deformations of metrics associated to deformations of complex structures. Let $\Sigma_z$ be a Riemann surface with local holomorphic coordinates~$z$, and let~$\mu$ be a Beltrami dif\/ferential which deforms the complex structure $z$ to $w$, for the  Riemann surface~$\Sigma_w$. Let the metrics corresponding to $\Sigma _z$ and $\Sigma _w$ be denoted respectively by $ds^2= 2 g_{\bar zz} dzd\bar z$ and  $d\tilde s^2= 2 \tilde g_{\bar ww} dw d\bar w$. The covariant derivatives $\nabla^z$ on a form $\varphi=\varphi(z) (dz)^n$ in $T^{n,0}(\Sigma_z)$ and $\tilde\nabla^w$ on a form $\tilde \varphi= \tilde \varphi (w) (dw)^n$ in  $T^{n,0}(\Sigma_w)$ are def\/ined as usual by relations which do not require a~connection
\begin{gather*}
\nabla^z\varphi   =   g^{z\bar z}   \partial_{\bar z}\varphi,
\qquad
\tilde\nabla^w\tilde\varphi   =   \tilde g^{w\bar w}   \partial_{\bar w}\tilde\varphi.
\end{gather*}
We now def\/ine the deformation $\tilde\nabla^z$ of the covariant derivative $\nabla^z$ by the Beltrami dif\/ferential~$\mu$ to be the following operator acting on a form $\varphi (z) (dz)^n$ in $T^{n,0}(\Sigma_z)$
\begin{gather*}
\tilde\nabla^z\varphi
=
\iota_{w\leftarrow z}^{-1}(\tilde\nabla^w(\iota_{w\leftarrow z}\varphi)).
\end{gather*}
The following surprisingly simple and exact formula for $\tilde\nabla^z$ holds to all orders in~$\mu$.
With the help of the  Weyl factor $e^{2\sigma}=|\partial_z w|^2 g^{z\bar z}\tilde g_{w\bar w}$ between the metrics, one f\/inds
\begin{gather}
\label{tildenabla}
\tilde\nabla^z\varphi
=
{e^{-2\sigma}  g^{z\bar z} \over
(1-\mu\bar\mu)^2}
\big(\partial_{\bar z}\varphi+\mu \partial_z\varphi
+n \varphi  \partial_z\mu
-n\varphi(\partial_{\bar z}+\mu\partial_z) \ln (1-\mu\bar\mu)\big),
\end{gather}
where $\mu$, $\bar \mu$, and $\varphi$ are functions of $z$. To the best of the authors' knowledge, this formula is new. It
is a generalization to all orders in $\mu$, namely for a f\/inite deformation of complex structures, of the well-known formulas of Friedan~\cite{Friedan} for the case of inf\/initesimal~$\mu$. Finite deformations of Abelian dif\/ferentials on Riemann surfaces have also been considered in the mathematical literature, in particular in~\cite{R,Y}.

To establish (\ref{tildenabla}), we start from the def\/ining formula for $\tilde\nabla^z$
\begin{gather*}
\tilde \nabla^z\varphi =
(\partial_w z)^{-(n-1)}\tilde g^{w\bar w} \, \partial_{\bar w} \big ( \varphi (z)   (\partial_w z)^n \big ).
\end{gather*}
Expressing  $\partial_{\bar w}$ in terms of derivatives with respect to $z$ and $\bar z$ using the second line in
(\ref{2a4}), and eliminating the metric $\tilde g^{w\bar w}$ using the def\/inition of the Weyl factor, we f\/ind
\begin{gather*}
\tilde \nabla^z\varphi = { e^{-2\sigma}   g^{z \bar z} \over (1-\mu \bar \mu)^2} \big (
(\partial_{\bar z}+\mu(z)\partial_z) \varphi (z) + n \varphi(z) (\partial_{\bar z}+\mu(z)\partial_z) \ln (\partial_w z) \big ).
\end{gather*}
Since all functions are now with respect to $z$,  we shall no longer exhibit this dependence below.
Using (\ref{3a3}) to eliminate $\ln (\partial_w z) $ in terms of $\ln (\partial_z w)$ and $\ln (1-\mu \bar \mu)$, we f\/ind
\begin{gather*}
\tilde \nabla^z\varphi = { e^{-2\sigma} \, g^{z \bar z} \over (1-\mu \bar \mu)^2} \big (
(\partial_{\bar z}+\mu \partial_z) \varphi  - n \varphi (\partial_{\bar z}+\mu \partial_z) \big \{ \ln (\partial_z w) + \ln (1-\mu \bar \mu) \big \} \big ).
\end{gather*}
It is straightforward to establish the following relation
\begin{gather*}
\partial_z \mu + (\partial_{\bar z}+\mu \partial_z) \ln (\partial_z w) =0
\end{gather*}
with the help of which the term in $\ln (\partial_z w)$ may be eliminated to give~(\ref{tildenabla}).
The derivation of~(\ref{tildenabla}) is complete.
The formula (\ref{tildenabla}) can be yet recast in a perhaps more suggestive form
\begin{gather*}
\tilde\nabla^z\varphi
={e^{-2\sigma}\over
(1-\mu\bar\mu)^{2-n}}
\hat\nabla^z\hat\varphi,
\end{gather*}
where we have set $\hat\varphi=(1-\mu\bar\mu)^{-n}\varphi$ and
\begin{gather}
\label{3c2}
\hat \nabla^z \hat \varphi = g^{z\bar z} \big ( \partial_{\bar z} \hat \varphi + \mu\partial_z\hat\varphi+n(\partial_z\mu)\hat\varphi\big).
\end{gather}
The covariant derivative $\hat \nabla ^z \hat \varphi$ now precisely coincides with the covariant derivative to f\/irst order in~$\mu$, though~(\ref{3c2}) is now valid for f\/inite deformations~$\mu$.

\section{Finite deformations via the stress tensor}

Using the theory and practice of f\/inite deformations of complex structures developed in the preceding sections, we shall now apply these f\/inite deformations to the standard worldsheet actions for superstring theory, namely for the scalar matter f\/ield~$x^\nu $ and the spinor matter f\/ield~$\psi^\nu$ in f\/lat space-time, and for the ghost  f\/ields $b$, $c$, $\beta$, $\gamma$. (Actions for non-linear $\sigma$-models for string theory in curved space-time may be treated in an analogous  manner.) We will recover the same actions that one would obtain by including a f\/inite deformation of the worldsheet metric, as expected. We will then show that these f\/inite deformations are fully accounted for by the inclusion of the usual stress tensors for these f\/ields, provided the overall total central charge vanishes.

\subsection{Finite deformations of worldsheet actions}

The simplest case is the action for the scalar f\/ields~$x^\nu $. Denoting their deformations to f\/ields in $T^{(0,0)}(\Sigma _w)$ by $\tilde x^\nu$, the scalar action is given by
\begin{gather*}
\tilde I_x= { 1 \over 4 \pi} \int_\Sigma d^2w \,\partial_w \tilde x^\nu  \partial_{\bar w} \tilde x^\nu,
\end{gather*}
where $\nu$ is the 10-dimensional space-time index, contracted with the help of the f\/lat Minkowski space-time metric by summation over repeated indices $\nu$. Using the f\/irst equation in~(\ref{2a4}) to recast the measure in terms of $z$, the second equation of~(\ref{2a4}) to recast the derivatives in terms of derivatives with respect to~$z$, and denoting the corresponding scalar f\/ield in $T^{(0,0)} (\Sigma _z)$ by $x^\nu$, we obtain the following action
\begin{gather*}
\tilde I_x
=
{ 1 \over 4 \pi} \int {d^2z\over 1-\mu\bar\mu}   \big( (1+\mu\bar\mu) \partial_z x^\nu \partial_{\bar z} x^\nu
+
\bar\mu   \partial_{\bar z}x ^\nu \partial_{\bar z}x ^\nu  + \mu   \partial_z x^\nu \partial_z x^\nu  \big).
\end{gather*}
Clearly, this action is not new; one arrives at the same expression for $\tilde I_x$  by considering the action for the scalar f\/ield $x^\nu$
\begin{gather*}
I_x = { 1 \over 8 \pi} \int _\Sigma d^2 \xi \sqrt{g} g^{mn} \partial_m x^\nu \partial_n x^\nu
\end{gather*}
in the presence of a general worldsheet metric~$g_{mn}$ with $m,n=1,2$, for arbitrary coordinates~$\xi^1$,~$\xi^2$ on~$\Sigma$. To recover $\tilde I_x$, it then suf\/f\/ices to choose the following parametrization
\begin{gather}
\label{4b4}
ds^2 = g_{mn} d\xi^m d \xi^n = 2   e^{2 \sigma (z)} |dz - \mu (z) d\bar z|^2.
\end{gather}
The dependence on  $ \sigma (z)$ again drops out in view of the Weyl invariance of the action~$I_x$.

The  action for the spin 1/2 f\/ield $\tilde \psi ^\nu$ belonging to $ T^{(\frac{1}{2}, 0)}(\Sigma _w)$ is given by
\begin{gather*}
\tilde I_\psi = - { 1 \over 4 \pi} \int _\Sigma d^2w \, \tilde \psi ^\nu \partial_{\bar w} \tilde \psi ^\nu.
\end{gather*}
In terms of the f\/ield $\psi ^\nu \in T^{(\frac{1}{2}, 0)} (\Sigma _z)$, it is given by
\begin{gather*}
\tilde I_\psi = - { 1 \over 4 \pi} \int _\Sigma { d^2 z \over 1 - \mu \bar \mu}  \big (
\psi ^\nu \partial_{\bar z} \psi ^\nu + \mu   \psi ^\nu \partial_z \psi ^\nu  \big ).
\end{gather*}
The contribution arising from terms of the form $\psi ^\nu \psi ^\nu$ actually vanishes in view of the Grassmann nature of the f\/ield~$\psi^\nu$. The resulting action in turn may be obtained alternatively by starting from an action for~$\psi ^\nu$ for a general metric parametrized by~(\ref{4b4}).

{\sloppy Finally, we consider the theory of pairs of f\/ields $b(z)$, $c(z)$ belonging to $T^{n,0}(\Sigma_z)$ and $T^{1-n,0}(\Sigma_z)$ respectively. Denote by $\tilde b (w)$ and $\tilde c(w)$ respectively their deformations to f\/ields in $T^{n,0}(\Sigma_w)$ and $T^{1-n,0}(\Sigma_w)$. The action is given by
\begin{gather*}
\tilde I_{bc} = { 1 \over 2 \pi}  \int d^2w \, \tilde b  \partial_{\bar w} \tilde c.
\end{gather*}
In terms of the f\/ields $b(z)$, $c(z)$, the deformed action is given by
\begin{gather*}
\tilde I_{bc} = { 1 \over 2 \pi} \int _\Sigma {d^2 z \over 1 -\mu \bar \mu}
\big ( b \partial_{\bar z} c + \mu b \partial_z c -(n-1) bc \partial_z \mu +(n-1) bc (\partial_{\bar z} + \mu \partial_z ) \ln (1-\mu \bar \mu) \big ).
\end{gather*}
The action for the superghosts $\beta$, $\gamma$ is obtained by replacing~$b$ by~$\beta$ and~$c$ by~$\gamma$ and setting~$n={3 \over 2}$.
For the case of $n={1\over 2}$ the~$b$ and~$c$ f\/ields have the same weight. The case of the $\psi$ f\/ield
treated earlier then corresponds to identifying~$b$ and~$c$ which is possible only at~$n=\frac{1}{2}$.

}

Putting together the cases of $n=0$, $n={1\over 2}$, $n=2$, and $n={3\over 2}$, we will obtain the expression to all orders in the deformation~$\mu$  for the worldsheet action for the RNS  string, at vanishing worldsheet gravitino f\/ield. It is well-known how to include the latter contribution.

\subsection{Chiral splitting}

Each classical action, $\tilde I_x$, $\tilde I_\psi$, $\tilde I_{bc}$, and $\tilde I_{\beta\gamma}$, depends on both~$\mu$ and~$\bar \mu$.
It is a basic principle of two dimensional conformal f\/ield theories that, in the deformation of their correlation functions, all terms involving $\mu\bar\mu$ should appear with a coef\/f\/icient proportional to their central charges.  Indeed, each quantum partition function depends on both~$\mu$ and~$\bar \mu$, but it is well-known from Belavin and Knizhnik~\cite{Belavin:1986cy} that this mixed dependence is proportional to the central charge of each f\/ield. The quantum partition function of the combined f\/ields $x^\nu$, $\psi ^\nu$, $b$, $c$, $\bar b$, $\bar c$, $\beta$, $\gamma$, $\bar \beta$, $\bar \gamma $ has vanishing central charge and, properly normalized, is the absolute value square of a chiral partition function, which only depends on~$\mu$ and not on~$\bar \mu$. A generalization of this result to the case of the full worldsheet supergravity including  the dependence on the Beltrami dif\/ferential~$\mu$ and the worldsheet gravitino f\/ield~$\chi$ was proven in~\cite{D'Hoker:1986bg}.

The above result may be further generalized to the case of superstring amplitudes with~$N$ external string states represented by vertex operator insertions on the worldsheet, and gives rise to the {\it chiral splitting theorem}, proven in~\cite{D'Hoker:1989ai} for strings propagating in f\/lat Minkowski space-time or on a f\/lat toroidal compactif\/ication thereof. In brief, the theorem states that a string amplitude in which left and right movers are complex conjugates of one another, with equal f\/ixed internal loop momenta $p$ and equal f\/ixed spin structure $\delta$, is the absolute value square of a~{\it chiral string amplitude} which depends only on $\mu$ and $\chi$ but not on their complex conjugates. The physical Type II string amplitudes are then obtained by pairing left and right chiral amplitudes at the same internal loop momenta $p$, but dif\/ferent spin structures~$\delta _L$ and~$\delta_R$ which are to be summed over independently  in accord with the GSO projection. For the heterotic strings, the prescription is analogous with the right moving chiral amplitude replaced by the chiral half of the bosonic string compactif\/ied on a 16-dimensional torus. In summary, the basic building blocks in the perturbation theory of all closed oriented string theories are the chiral string amplitudes which depend on $\mu$ but are independent of~$\bar \mu$.

\subsection{Finite deformations via the stress tensor}

Thus, in a theory with total central charge $0$, as in superstring perturbation theory, we can just drop all the dependence involving $\mu\bar\mu$.  In view of all the formulas derived in the previous sections for f\/inite deformations of complex structures, we f\/ind the remarkable fact that the covariant derivatives, the actions, and the stress tensors are all given by the same formulas as for inf\/initesimal deformations, but with the property that the formulas are valid to all orders of expansion in $\mu$. The corresponding ef\/fective chiral action for the chiral parts of the f\/ields (denoted $x_+^\nu$ and $\psi _+^\nu$ for the matter RNS f\/ields) of the superstring thus takes the form
\begin{gather*}
I = {1 \over 2 \pi} \int _\Sigma d^2z
\left( \frac{1}{2} \partial_{\bar z} x^\nu_+ \partial_z x^\nu_+ - \frac{1}{2} \psi _+^\nu \partial_{\bar z} \psi _+ ^\nu
+ b \partial_{\bar z} c + \beta \partial_{\bar z} \gamma - \mu T - \chi S \right),
\end{gather*}
where $T$ is the worldsheet stress tensor obtained from the expressions of the actions given above, and truncated by setting $\bar \mu=0$. Also,  $S$ is the worldsheet supercurrent, whose ef\/fects have been included for completeness, but without derivation. This formula was the starting point for the construction of two-loop amplitudes with NS vertex operators in \cite{D'Hoker:2001zp,D'Hoker:2005}, where the formula was needed only to f\/irst order in~$\mu$, but here it is established to all orders in~$\mu$. We also refer to~\cite[Section 3.5]{Witten:2012ga} for a helpful treatment of deformations of supercomplex structures by f\/ields.

Collecting all terms for the stress tensor, we f\/ind the well-known form \cite{Friedan:1985ge}
\begin{gather*}
T = - \frac{1}{2} \partial_z x^\nu _+ \partial_z x^\nu _+ + \frac{1}{2} \psi ^\nu \partial_z \psi _+^\nu  - b \partial_z c - \partial_z(bc)
- \beta \partial_z \gamma  - \frac{1}{2} \partial_z(\beta \gamma ).
\end{gather*}
The expression for the supercurrent is similarly given by \cite{Friedan:1985ge}
\begin{gather*}
S = - \frac{1}{2} \psi ^\nu _+ \partial_z x_+^\nu + \frac{1}{2} b \gamma -\frac{1}{2} \beta \partial_z c - \partial_z (\beta c).
\end{gather*}
Note that neither $T$ nor $S$ involves any dependence on~$\mu$, a key ingredient in the conclusion that the formulas for deformations by~$\mu$ customarily viewed as valid only to f\/irst order in~$\mu$, are actually valid to all orders in $\mu$. Also note that~$\chi$ is independent of~$\mu$, so that when  the above action is used  to higher order in~$\mu$, there is no need to deform the complex structure in which~$\chi$ was originally def\/ined: all such deformations will ef\/fectively be taken care of by the insertion of the stress tensor term~$\mu T$.

\subsection{Finite deformations of Green functions}

The chiral splitting principle brings about an enormous simplif\/ication. We shall now test this explicitly
by showing that the Green functions for the scalar f\/ield~$x^\nu_+$, for the spinor f\/ield~$\psi ^\nu_+$, and for the ghost systems~$b$,~$c$ and~$\beta$,~$\gamma$ are precisely given by the correlators for these f\/ields in the presence of their respective stress tensor, and this to all orders in~$\mu$. For the sake of clarity we shall treat each system  of f\/ields separately. Our main concern is with higher loop superstring amplitudes for closed oriented superstrings, so we shall discuss the case when~$\Sigma$ is a~closed oriented Riemann surface of genus~$h \geq 2$. For the case of the sphere $h=0$, the complex structure is unique, while for the torus~$h=1$ the situation is  well-known.

\subsection[The  $b$, $c$ system]{The  $\boldsymbol{b}$, $\boldsymbol{c}$ system}

The $b,c$ system has anti-commuting  f\/ields $b$, $c$ and will be considered here for arbitrary $n \geq {3 \over 2}$ (we shall discuss the cases of $n=0,\frac{1}{2}, 1$ separately below).  We shall show that the Green function for the Cauchy--Riemann operator $\hat \nabla ^z _\mu$ is identical to the correlator for the~$b$,~$c$~system in the presence of the stress tensor insertion $\mu T$, to all orders in~$\mu$.
Recall that the covariant derivative $\hat \nabla ^z _\mu$, def\/ined earlier in~(\ref{3c2}),
acting on a tensor f\/ield of weight $(n,0)$, is given by
\begin{gather}
\label{4d2}
\hat \nabla _\mu ^z = { 1 \over \sqrt{\tilde g} } \big ( \partial_{\bar z} + \mu \partial_z + n (\partial_z \mu) \big ).
\end{gather}
For the case considered here, namely $n \geq 3/2$, the kernel of the adjoint of $\hat \nabla ^z _\mu$ vanishes, and the Green function is simply def\/ined by
\begin{gather*}
%\label{4d3}
\tilde \nabla _\mu ^z G_\mu (z,z') = 2 \pi \delta _{\tilde g} (z,z').
\end{gather*}
By conformal invariance, all factors of $\tilde g$ cancel on both sides above  so that
the equations which def\/ine the Green functions for $\mu$ and for $\mu=0$ are given by
\begin{gather}
\left ( \partial_{\bar z} + \mu \partial_z + n (\partial_z \mu) \right ) G_\mu (z,z')  =  2 \pi \delta (z-z')
\nonumber \\
\partial_{\bar z} G_0 (z,z')  =  2 \pi \delta (z-z').\label{4d4}
\end{gather}
To  calculate $G_\mu$ in terms of $G_0$ to all orders in $\mu$, we combine the above equations
\begin{gather*}
%\label{4d5}
\partial_{\bar z} G_\mu (z,z') = \partial_{\bar z } G_0 (z,z') - \big ( \mu \partial_z + n (\partial_z \mu) \big )
G_\mu (z,z'),
\end{gather*}
which may be integrated as follows
\begin{gather}
\label{4d6}
G_\mu (z,z') =  G_0(z,z') - { 1 \over 2 \pi} \int d^2 v \, G_0(z,v)  \big ( \mu \partial_v + n (\partial_v \mu) \big )G _\mu (v,z').
\end{gather}
In general, this integration will allow for an additional contribution which is a holomorphic form of weight $(n,0)$ in $z$, since the  $\partial_{\bar z}$ operator on $(n,0)$-forms has a non-trivial kernel. This arbitrariness may be f\/ixed, for example, by insisting that the free zeros of $G_\mu$ in its f\/irst argument,~$z$, be independent of~$\mu$, and coincide with the free zeros of~$G_0$. Denoting these free zeros by $z_a$ with $a=1, \cdots, \Upsilon(n)$ where $\Upsilon (n)= (2n-1)(h-1)$, for the case $h \geq 2$ of interest here, we are led to require
\begin{gather}
\label{4d7}
G_\mu (z_a, z')=0,\qquad a = 1, \dots , \Upsilon (n),
\end{gather}
for all points  $z'$, arbitrary $\mu$ including 0, and~$z_a$ independent of~$\mu$. This requirement is natural in superstring perturbation theory. The Green function $G_\mu$ may be obtained to arbitrary order in~$\mu$ by iterating the integral  equation, and we get schematically
\begin{gather}
\label{4d8}
G_\mu = G_0 + G_0 M G_0 + G_0 M G_0 M G_0 + \cdots,
\end{gather}
where $M$ is the operator $ - ( \mu \partial_v + n (\partial_v \mu)  )/(2\pi)$.

\subsubsection{Green function by deformed correlator}

We shall now prove that the Green function $G_\mu$ is alternatively given in terms of the chiral deformed correlator for the $bc$ system by
\begin{gather}
\label{4d9}
G_\mu (z, z' ) = \langle  b(z) c(z') \rangle  _\mu,
\end{gather}
where the correlator $\langle  b(z) c(z') \rangle  _\mu$  is def\/ined as follows
\begin{gather*}
\langle  b(z)  c(z') \rangle  _\mu = { 1 \over Z_{bc} } \left \langle  b(z) c(z') \prod _{a=1}^{\Upsilon (n)} b(z_a)  \exp \left \{
 {1 \over 2\pi} \int d^2 v \, \mu(v)  T_{bc} (v) )  \right \} \right \rangle _0
\end{gather*}
and the stress tensor is given by
\begin{gather}
\label{Tbc}
T _{bc} = - b \partial_z c - (n-1) \partial_z (bc).
\end{gather}
The normalization factor $Z_{bc} $ is def\/ined by omitting the insertions $b(z)c(z')$, and is given by
\begin{gather*}
%\label{4d10}
 Z_{bc}  = \left \langle   \prod _{a=1}^{\Upsilon (n)} b(z_a) \, \exp \left \{
 {1 \over 2\pi} \int d^2 v \, \mu(v)  T_{bc} (v) )  \right \} \right \rangle _0.
 \end{gather*}
The expectation value $\langle  \cdots \rangle _0$ is with respect to the action for the $bc$ system
for vanishing $\mu$. In particular, $\langle  b (z) c(z') \rangle _0$ equals $G_0(z,z')$ and has a pole of unit residue at $z=z'$
\begin{gather}
\label{resid}
\langle  b(z)  c(z') \rangle  _0 = G_0(z,z')= { 1 \over z-z'} + {\mathcal O}\big( ( z-z')^0\big).
\end{gather}
To prove the proposed equality (\ref{4d9}), one may proceed either by showing that the above expression obeys the dif\/ferential equation for~$G_\mu$ in~(\ref{4d4}) as well as the normalization conditions~(\ref{4d7}), or alternatively that its expansion in powers of~$\mu$ coincides with the one given in~(\ref{4d8}). It is instructive to give both proofs.

By construction, both $G_\mu (z,z')$ and $\langle  b(z)   c(z') \rangle  _\mu$ are forms of weight $(n,0)$ in $z$ and weight $(1-n,0)$ in $z'$, both vanish at $z=z_a$ for all $a=1,\dots, \Upsilon (n)$, and both have  a simple pole with unit residue in~$z$ at $z=z'$. Furthermore, the operator product of the f\/ields $b$ and $T$ exhibits a~universal pole governed by the dimension~$n$ of~$b$, namely,
\begin{gather*}
b(z)  T_{bc}(v)  = { n   b(v) \over (v-z)^2}  + (n-1) { \partial_v b(v) \over z-v} + {\mathcal O}\big( ( z-v)^0\big).
\end{gather*}
Upon integrating $T_{bc}$ against $\mu$, and applying the Cauchy--Riemann operator $\partial_{\bar z}$, we have
\begin{gather*}
\partial_{\bar z} \left ( b(z) \int _\Sigma d^2 v \, \mu(v) T_{bc} (v) \right ) = - \mu(z) \partial_z b(z) - n (\partial_z \mu(z) )  b(z).
\end{gather*}
As a result, one f\/inds that $\langle  b(z)   c(z') \rangle  _\mu$ obeys the f\/irst equation in~(\ref{4d4}).
Since it also obeys the normalization conditions~(\ref{4d7}) this completes the proof of~(\ref{4d9}) to all orders in~$\mu$.

Alternatively, one may compare the expansions of $G_\mu(z,z')$ and $\langle  b(z) \, c(z') \rangle  _\mu$  in powers of $\mu$ term by term.  Concentrating, for example,  on the term of f\/irst order in $\mu$, we have
\begin{gather*}
%\label{4d11}
\langle  b(z) c(z') \rangle  _\mu = \langle  b(z) c(z') \rangle  _0
+ {1 \over 2\pi} \int d^2 v \, \mu(v)   \langle  T_{bc} (v)   b(z)   c(z') \rangle _0 + {\mathcal O}\big(\mu^2\big),
\end{gather*}
where $T$ was given in~(\ref{Tbc}). Using Wick contractions to carry out the correlator, we f\/ind
\begin{gather}
\int d^2 v \, \mu(v)   \langle  T_{bc} (v)  b(z)   c(z') \rangle _0
 = - \int _\Sigma d^2 v \, G_0(z,v) \big  ( \mu(v) \partial_v + n \partial_v \mu(v) \big ) G_0(v,z').
 \end{gather}
Comparison with (\ref{4d6}) shows perfect agreement. Terms of higher order may be identif\/ied in an analogous manner.

\subsection[The $\beta$, $\gamma$ system]{The $\boldsymbol{\beta}$, $\boldsymbol{\gamma}$ system}

The only dif\/ference between the  $\beta$, $\gamma$ system and the~$b$,~$c$ system discussed above is that~$\beta$ and~$\gamma$ are commuting f\/ields while~$b$,~$c$ were anti-commuting f\/ields. The~$\beta$,~$\gamma$ system will be considered here for~$n \geq { 3 \over 2}$. In particular, the Green function  for the Cauchy--Riemann operator~$\tilde \nabla ^z_\mu$ is identical to the one for the $b$, $c$ system, and is therefore def\/ined by the dif\/ferential equation~(\ref{4d4}) along with the normalization conditions (\ref{4d7}) for the appropriate value of~$n$, namely $n={3 \over 2} $ in the case of the superstring superghost system. The result for the~$\beta$,~$\gamma$ system is as follows
\begin{gather}
\label{4d12}
G_\mu (z, z' ) = - \langle  \beta (z) \gamma (z') \rangle  _\mu,
\end{gather}
where the correlator $\langle  \beta (z) \gamma (z') \rangle  _\mu$  is def\/ined as follows
\begin{gather*}
\langle  \beta (z)   \gamma (z') \rangle  _\mu
= { 1 \over Z_{\beta \gamma} } \left \langle  \beta (z) \gamma (z') \prod _{a=1}^{\Upsilon (n)}
\delta \big ( \beta (z_a) \big )  \exp \left \{
 {1 \over 2\pi} \int d^2 v \, \mu(v)    T_{\beta \gamma } (v) )  \right \} \right \rangle _0
\end{gather*}
and the stress tensor is given by
\begin{gather*}
%\label{4d13}
T _{\beta \gamma} = - \beta \partial_z \gamma - (n-1) \partial_z (\beta \gamma ).
\end{gather*}
The normalization factor $Z_{\beta \gamma} $ is def\/ined by omitting the insertions $\beta (z) \gamma (z')$, and  given by
\begin{gather*}
%\label{4d14}
 Z_{\beta \gamma}  = \left \langle   \prod _{a=1}^{\Upsilon (n)} \delta \big ( \beta (z_a) \big )   \exp \left \{
 {1 \over 2\pi} \int d^2 v \, \mu(v)   T_{\beta \gamma } (v) )  \right \} \right \rangle _0.
 \end{gather*}
Note that the sign dif\/ference in~(\ref{4d12}) constitutes a crucial dif\/ference with the~$b$,~$c$ system.
In particular, the correlator at~$\mu=0$ has a single pole at $z=z'$ with residue~$-1$, as given by~(\ref{resid}).
For  useful treatments of the~$\beta$,~$\gamma$ correlators, we refer the reader to~\cite{Lechtenfeld:1989wu,Verlinde:1986kw} and  \cite[Section~10]{Witten:2012bh}.

\subsection[The spinor f\/ield $\psi ^\mu_+$]{The spinor f\/ield $\boldsymbol{\psi ^\mu_+}$}

The spinor f\/ield $\psi$ (we shall drop the Lorentz superscript $\nu$ and the chirality subscript~$+$ in this section), may be viewed as the  special case of the $b,c$ system with $n=\frac{1}{2}$ and the f\/ields $b$ and $c$  identif\/ied. The Green function is now referred to as the Szeg\"o kernel. The Cauchy--Riemann operator $\tilde \nabla ^z _\mu$  is given by~(\ref{4d2}) with $n=\frac{1}{2}$. For even spin structures, the kernel of~$\tilde \nabla ^z _\mu$ and the kernel of its adjoint operator are both null, generically on moduli space. For odd spin structure, however, both operators have a one-dimensional kernel, again generically on moduli space.

For even spin structure, the Szeg\"o kernel  will be denoted by $S_\mu (z, z')$. It is odd under the interchange of $z$ and $z'$ and  satisf\/ies the dif\/ferential equation
\begin{gather*}
\left ( \partial_{\bar z} + \mu \partial_z + \frac{1}{2} (\partial_z \mu) \right ) S_\mu (z,z')  =  2 \pi \delta (z-z')
\nonumber \\
\partial_{\bar z}  S_0 (z,z')  =  2 \pi \delta (z-z')%\label{4f1}
\end{gather*}
away from non-generic points on moduli space where the kernel and co-kernel of $\tilde \nabla ^z _\mu$ are non-trivial.  The expression for $S_\mu (z, z')$ in terms of $S_0$ and $\mu$ to all orders in $\mu$ is given by an immediate adaptation of (\ref{4d6}) to this case, namely by iterating to an arbitrary order in $\mu$ the integral equation
\begin{gather*}
%\label{4f2}
S_\mu (z,z')
=  S_0(z,z') - { 1 \over 2 \pi} \int d^2 v \, S_0(z,v)  \left  ( \mu \partial_v + \frac{1}{2}  (\partial_v \mu) \right )S_\mu (v,z').
\end{gather*}
The deformed Szeg\"o kernel is then related to the deformed correlator of the f\/ield $\psi$ by
\begin{gather*}
%\label{4f3}
S_\mu (z, z' ) = - \langle  \psi (z) \psi (z') \rangle  _\mu,
\end{gather*}
where the correlator $\langle  \psi (z) \psi (z') \rangle  _\mu$  is def\/ined as follows,
\begin{gather*}
\langle  \psi (z)   \psi (z') \rangle  _\mu
= { 1 \over Z_{\psi} } \left \langle  \psi (z) \psi (z')  \exp \left \{
 {1 \over 2\pi} \int d^2 v \, \mu(v)  T_{\psi} (v) )  \right \} \right \rangle _0
\end{gather*}
and the stress tensor is given by
\begin{gather*}
%\label{Tpsi}
T _{\psi} =  \frac{1}{2} \psi  \partial_z \psi.
\end{gather*}
The normalization factor $Z_{\psi} $ is def\/ined by omitting the insertions $\psi (z) \psi (z')$, and is given by{\samepage
\begin{gather*}
%\label{4f5}
 Z_{\psi}  = \left \langle    \exp \left \{
 {1 \over 2\pi} \int d^2 v \, \mu(v)  T_{\psi} (v) )  \right \} \right \rangle _0.
 \end{gather*}
The expectation value $\langle  \cdots \rangle _0$ is with respect to the action for the $\psi$ system
for vanishing~$\mu$.}

For odd spin structure, generically on moduli space, the Cauchy--Riemann operator $\tilde \nabla _\mu ^z$, as well as its adjoint, have one zero mode which we shall denote by $h_\mu(z)$, and which satisf\/ies
\begin{gather}
\label{4f6}
\partial_{\bar z } h_\mu + \mu \partial_z h _\mu + \frac{1}{2} (\partial_z \mu) h_\mu =0.
\end{gather}
Due to the presence of this zero mode, there is no unique way of def\/ining the Szeg\"o kernel,
and it is generally not odd under the interchange of~$z$,~$z'$. To make the connection with correlators and chiral splitting, it will be convenient to consider a pair of chiral fermions~$\psi ^1$,~$\psi^2$, and denote their complex linear combinations as follows, $\psi =(\psi ^1 + i \psi ^2)/\sqrt{2}$ and $\bar \psi= (\psi ^1 - i \psi ^2)/\sqrt{2}$. One may think of $\psi$ and $ \bar \psi$ as the f\/ields $b$ and $c$ of the $n=\frac{1}{2}$ system {\it without having made the identification of the field~$b$ with the field~$c$}. Introducing on $\Sigma$ an arbitrary point $\zeta$  at which we have $h_\mu (\zeta) \not=0$,  the Szeg\"o kernel~$S_\mu (z,z'; \zeta) $ may be def\/ined as satisfying the following dif\/ferential equation
\begin{gather}
\label{4f7}
\left ( \partial_{\bar z} + \mu \partial_z + \frac{1}{2} (\partial_z \mu) \right ) S_\mu (z,z'; \zeta )  =  2 \pi \delta (z-z') - 2 \pi \delta (z-\zeta)   { h _\mu (z') \over h_\mu (\zeta)}.
\end{gather}
Integration of both sides against $h_\mu (z)$ vanishes, and this is the rationale for the choice of the subtraction term on the right hand side. When $z'=\zeta$, the right hand side vanishes, so that $S_\mu (z, z' ; \zeta ) |_{z'=\zeta}$ must be proportional to $h_\mu (z)$. Therefore, it is natural to set the Szeg\"o kernel to zero at $z'=\zeta$. This condition is analogous to (\ref{4d7}) for the $b,c$-system, and  def\/ines the Szeg\"o kernel uniquely. We shall denote the unique free zero in $z$ of $S_\mu (z,z';\zeta)$ by~$\zeta'$.

We are led to proposing the following identif\/ication of the Szeg\"o and correlator
\begin{gather}
\label{4f8}
S_\mu (z,z';\zeta) = - \langle  \psi (z) \bar \psi (z') \rangle  _\mu,
\end{gather}
where the correlator is given by
\begin{gather}
\label{corr}
\langle  \psi (z)   \bar \psi (z') \rangle  _\mu = { 1 \over Z_{\psi \bar \psi } } \left \langle  \psi (z)  \bar \psi (z') \psi (\zeta ')  \bar \psi (\zeta )  \exp \left \{
 {1 \over 2\pi} \int d^2 v \, \mu(v)   T_{\psi \bar \psi } (v) )  \right \} \right \rangle _0
\end{gather}
and the stress tensor is given by
\begin{gather*}
%\label{Tbc}
T _{\psi \bar \psi } = - \frac{1}{2} \psi    \partial_z \bar \psi  +\frac{1}{2}  \partial_z \psi   \bar \psi .
\end{gather*}
The normalization factor $Z_{\psi \bar \psi } $ is def\/ined by omitting the insertions $\psi (z) \bar \psi (z')$, and given by
\begin{gather*}
%\label{4d10}
 Z_{\psi \bar \psi }  = \left \langle   \psi (\zeta ')   \bar \psi (\zeta )   \exp \left \{
 {1 \over 2\pi} \int d^2 v \, \mu(v)   T_{\psi \bar \psi} (v) )  \right \} \right \rangle _0.
 \end{gather*}
The proposed relation~(\ref{4f8}) may be proven by applying the operator $\partial_{\bar z} + \mu \partial_z +\frac{1}{2} (\partial_z\mu)$ to the correlator (\ref{corr}) and using the OPE of $T_{\psi \bar \psi} $ with $\psi$. The normalization factor~$Z_{\psi \bar \psi}$ guarantees that the f\/irst~$\delta$-function on the right side of~(\ref{4f7}) arises with the proper factor of~$2 \pi$. The factor of the second term on the right side of~(\ref{4f7}) arises with the help of  the following  relation
\begin{gather*}
{h_\mu (z') \over h_\mu (\zeta)} =  { 1 \over Z_{\psi \bar \psi } } \left \langle   \psi (\zeta ') \bar \psi (z')    \exp \left \{
 {1 \over 2\pi} \int d^2 v \, \mu(v)   T_{\psi \bar \psi } (v) )  \right \} \right \rangle _0,
 \end{gather*}
which is easily proven by showing that the right hand side satisf\/ies~(\ref{4f6}) as a function of~$z'$,
and is equal to~1 at the point~$z'=\zeta$.

\subsection[The scalar f\/ield $x^\nu_+$]{The scalar f\/ield $\boldsymbol{x^\nu_+}$}

The scalar f\/ield $x$ (we shall omit the superscript $\nu$ and the subscript~$+$ in this section) is not properly a conformal f\/ield due to the presence of the constant zero mode, and it is preferable to work with its derivative $\partial_z x$ which is a conformal f\/ield of type~$(1,0)$. The corresponding Green function is the third Abelian dif\/ferential of the second kind, $\omega (z,z') = \partial_z \partial_{z'} \ln E(z,z')$ for vanishing $\mu$, and generalizes as follows to the case of arbitrary~$\mu$
\begin{gather*}
\partial_{\bar z}   \omega _\mu (z, z') + \partial_z \big ( \mu (z)   \omega _\mu (z,z') \big ) =  2 \pi \partial_{z'} \delta (z-z').
\end{gather*}
Note that both sides properly integrate to zero against a constant.
The identif\/ication with the chiral correlator of the f\/ield $x$ is as follows
\begin{gather*}
\omega_\mu  (z,z') = - \langle  \partial_z x (z')   \partial_z x(z') \rangle  _\mu,
\end{gather*}
which is consistent with the well-known result at $\mu=0$ given by
\begin{gather*}
\langle  x(z)   x(z') \rangle  _0 = - \ln E(z, z')
\end{gather*}
and its derivatives in $z$ and $z'$.
The correlator is def\/ined as follows
\begin{gather*}
\langle  \partial_z x (z')   \partial_z x(z') \rangle  _\mu = { 1 \over Z_x} \left \langle  \partial_z x (z)    \partial_z (z') \exp \left \{
 {1 \over 2\pi} \int d^2 v \, \mu(v)    T_{x } (v) )  \right \} \right \rangle _0.
\end{gather*}
The stress tensor is given by
\begin{gather*}
T_x = - \frac{1}{2} \partial_z x   \partial_z x
\end{gather*}
and the normalization factor is given by
\begin{gather*}
Z_x= \left \langle   \exp \left \{   {1 \over 2\pi} \int d^2 v \, \mu(v)    T_{x } (v) )  \right \} \right \rangle _0.
\end{gather*}
In both correlators, only the chiral part of the f\/ield enters.
This completes our derivation of the deformation to all orders in~$\mu$ of every conformal f\/ield theory needed in the critical superstring in f\/lat Minkowski space-time, and toroidal compactif\/ications thereof.

\subsection*{Acknowledgements}

We would like to acknowledge the organizers, Decio Levi, Willard Miller, Yvan Saint-Aubin, and Pavel Winternitz for inviting us to  the enjoyable conference and celebration in honor of Luc Vinet at the Centre de Recherches Math\'ematiques.
We would also like to thank the referees for their very careful reading of the paper.
This research was supported in part by the National Science Foundation grants PHY-13-13986 and DMS-12-66033.

\pdfbookmark[1]{References}{ref}
\LastPageEnding

\end{document}